\documentclass[a4paper]{article}

\usepackage{INTERSPEECH2019}
\usepackage{amsmath,graphicx,multirow,booktabs,float,threeparttable}
\newcommand{\tabincell}[2]{\begin{tabular}{@{}#1@{}}#2\end{tabular}}
\usepackage{lipsum}

\title{The DKU-SMIIP System for NIST 2018 Speaker Recognition Evaluation} 
\name{Danwei Cai$^1$, Weicheng Cai$^{1, 2}$, Ming Li$^1$\thanks{This research was funded in part by the National Natural Science Foundation of China (61773413), Natural Science Foundation of Guangzhou City
(201707010363), Six Talent Peaks project in Jiangsu Province (JY-074), Science and Technology Program of Guangzhou City (201903010040).}}
\address{
  $^1$Data Science Research Center, Duke Kunshan University, Kunshan, China\\
  $^2$School of Electronics and Information Technology, Sun Yat-sen University, Guangzhou, China}
\email{ming.li369@dukekunshan.edu.cn}

\begin{document}

\maketitle
\begin{abstract}
In this paper, we present the system submission for the NIST 2018 Speaker Recognition Evaluation by DKU Speech and Multi-Modal Intelligent Information Processing (SMIIP) Lab. We explore various kinds of state-of-the-art front-end extractors as well as back-end modeling for text-independent speaker verifications. Our submitted primary systems employ multiple state-of-the-art front-end extractors, including the MFCC i-vector,  the DNN tandem i-vector,  the TDNN x-vector, and the deep ResNet. After speaker embedding is extracted, we exploit several kinds of back-end modeling to perform variability compensation and domain adaptation for mismatch training and testing conditions. The final submitted system on the fixed condition obtains actual detection cost of 0.392 and 0.494 on CMN2 and VAST evaluation data respectively. After the official evaluation, we further extend our experiments by investigating multiple encoding layer designs and loss functions for the deep ResNet system. 

\end{abstract}

\section{Introduction}
Since 1996, the US National Institute of Standards and Technology (NIST) has been conducting speaker recognition evaluations (SRE) to explore promising new ideas and measure the performance the state-of-the-art speaker recognition systems~\cite{sadjadi_nist_2018}. NIST SRE 2018 focus on text-independent speaker verification~(TISV) and contains two testing tasks:

\begin{itemize}
    \item CMN2: Similar to SRE16, the development and evaluation data of CMN2 task are non-English conversational telephone speech collected with various telephone channels. The duration of the test segments ranges 10-60 seconds with enrollment limited to 1 minute. The channel-, language- and duration-mismatches impose great challenges to the CMN2 task. 
    \item VAST: the VAST task of SRE18 contains English audio extracted from YouTube videos that vary in duration from a few seconds to several minutes. The audios are obscured in complex environmental settings including reverberation and noises. Each audio recording may contain speech from multiple talkers. Manually produced diarization labels are provided for the enrollment audio but not for the test audio.
\end{itemize}

There are two major categories of methods to extract the speaker embedding for TISV. The first comprises stacking self-contained algorithmic components, and the representative is the classical i-vector approach~\cite{dehak_front-end_2011}. The i-vector extractor can be trained using either the acoustic-level features or phoneme discriminant features extracted from the additional acoustic model~\cite{Li2014SpeakerVA,li_generalized_2016}.

The second category relies on the model trained by a downstream procedure through a deep neural network~(DNN). The representative is the x-vector system~\cite{snyder_x-vectors:_2018} based on time delay neural network~(TDNN). Recently, Cai \textit{et al.} have achieved state-of-the-art performance by extracting speaker embeddings from a deep residual neural network~(ResNet)~\cite{cai_exploring_2018}. Based on the deep ResNet framework, various kinds of encoding layers like learnable dictionary encoding~(LDE) layer~\cite{cai_novel_2018} and NetVLAD layer~\cite{Cai_2018_ISCSLP} as well as loss functions like center loss~\cite{10.1007/978-3-319-46478-7_31, cai_exploring_2018} and angular softmax loss~(A-softmax)~\cite{liu_sphereface:_2017, cai_exploring_2018} are explored and improve the performance significantly. 

Our developed systems consist of multiple state-of-the-art front-end extractors, including MFCC i-vector,  DNN tandem i-vector,  TDNN x-vector, and deep ResNet. After speaker embeddings are extracted, we investigate several back-ends modeling and score normalization methods. Different components including variabilities compensation, domain adaptation, in-domain whitening and Gaussian Probability Linear Discriminant Analysis (PLDA) are used for scoring. Also, cosine similarity is used as an alternative scoring method for ResNet based systems.

This paper is organized as follows: Section 2 describes the details of our submitted system. Section 3 illustrates the submission performance of both the single and fusion systems. Some of our post-evaluation experiments and analysis are presented in section 4. Conclusions are drawn in section 5.

\section{System descriptions}

\subsection{Front-end extractor}

\subsubsection{MFCC i-vector}
The MFCC i-vector system is developed by adapting the Kaldi SRE16 recipe. 20-dimensional MFCC is augmented with their delta and double delta coefficients, making 60-dimensional feature vectors.  A simple energy-based voice activity detector (VAD) is used. A short-time cepstral mean subtraction~(CMS) is applied on the over a 3-second sliding window. A 2048-components full covariance Gaussian Mixture Model-Universal Background Model~(UBM) is trained, along with a 600-dimensional i-vector extractor \cite{dehak_front-end_2011}.

\subsubsection{DNN tandem i-vector}
The DNN tandem i-vector system uses the tandem feature, which has a phonetic-aware feature concatenated with the acoustic MFCC feature \cite{Li2014SpeakerVA, li_generalized_2016}. For phonetic feature extraction, we employed a DNN acoustic model, trained on Fisher dataset, with 5621 tied tri-phone states to get the frame-level phoneme posterior probability~(PPP). After logarithm and Principal Component Analysis (PCA), the resulted 52-dimensional phonetic aware features are fused with the 60-dimensional MFCC at the feature level to get the 112-dimensional tandem feature. After the tandem feature is extracted, the subsequent UBM and factor analysis setup is exactly the same as in the MFCC i-vector recipe. 

\subsubsection{TDNN x-vector}
The x-vector \cite{snyder_x-vectors:_2018} system is developed by adapting the Kaldi SRE16 recipe.  For the x-vector extractor, a DNN is trained to discriminate speakers in the training set. The first five timed delayed layers operate at frame-level. Then a temporal statistics pooling layer is employed to compute the mean and standard deviation over all frames for an input segment. The resulted segment-level representation is then fed into two fully connected layers to classify the speakers in the training set. After training, speaker embeddings are extracted from the 512-dimensional affine component of the first fully connected layer.

\subsubsection{Deep ResNet}
\label{front-resnet}

For feature extraction, each audio is converted to 64-dimensional log Mel-filterbank energies. CMS and VAD operation is performed the same as in the MFCC i-vector. 
 
%Concerning the fact that speech is translation invariant along the time-axis only, the x-vector is built rely on 1-dimensional convolutional layers and pooled only on time axis.  In contrast, our deep ResNet system using 2-dimensional convolutional layer. Consistent with our previous work~\cite{cai_novel_2018, cai_exploring_2018, Cai_2018_Interspeech},  The x-vector system employs 5-layer TDNN and the output channels are up to 1500. In contrast, although we have 34-layer ResNet, the number of output channels is only up to 256. In this sense,  our ResNet structure is much deeper but thinner.

We follow the deep ResNet system as described in~\cite{cai_exploring_2018,cai_insights_2018,Cai_2018_Interspeech}, and we increase the widths (number of channels) of the residual blocks from \{16, 32, 64, 128\} to \{32, 64, 128, 256\}. The network structure contains three main components: a front-end ResNet, a pooling layer, and a feed-forward network. Given a feature sequence of size $D \times L$, the ResNet learns three-dimensional feature descriptions of shape $C\times H\times W$, where $C$ denotes the number of channels, $H$ and $W$ denotes the height and width of the feature maps. To get the single utterance-level representation, we adopt a global average pooling (GAP) layer, which accumulates mean statistics for each feature map. Given feature maps $ \mathbf{F} \in \mathbb{R}^{C\times H\times W}$,  the output of GAP is defined as:
\begin{equation}
v_i =\frac{1}{H\times W}\sum_{j=1}^{j=H}\sum_{k=1}^{k=W} \mathbf{F}_{i,j,k}
\end{equation}
 
Therefore, we get fixed-dimensional utterance-level representation $\mathbf {V}={\left[v_1, v_2, \cdots, v_C\right]}$ for a variable-length utterance. We further process the utterance-level representation through a classifier with two fully-connected layers. In the output layer, each unit is represented as a target speaker identity. All the components in the pipeline are jointly learned in an end-to-end manner with a unified loss function.

We design a variable-length data loader to generate mini-batch training samples on the fly. For each training step, a random integer $L$ which indicates the frame-length is created. The data loader provides dynamic mini-batch data of shape $B \times D \times L$ before each training step, where $B$ denotes the batch size and $D$ denotes the feature dimension. Therefore, the length of the mini-batch training samples is a batch-wise variable number. 

After training, the 256-dimensional utterance-level speaker embedding is extracted after the penultimate layer of the neural network for the given utterance. In the testing stage, the full-length feature sequence is directly fed into the network, without any truncate or padding operation.  More detailed network setup and training config can refer to~\cite{cai_exploring_2018}.

\subsection{Back-end modeling}
We trained different back-ends for CMN2 task and VAST task. Details are described as follows.

\subsubsection{Variabilities compensation}
For variabilities compensation, we use either Linear Discriminant Analysis (LDA) or Locality Sensitive Discriminant Analysis (LSDA) \cite{cai2007locality, cai_locality_2016} to select the most speaker relevant feature subset and reduce the variabilities irrelevant to the speaker.

Before applying LDA or LSDA, the speaker discriminant features are mean centered. It is noticed that as the training data contains different dataset such as Voxceleb, Call My Net (CMN) and NIST SRE, we apply mean subtraction separately for each dataset to diminish the inter-dataset variabilities.

\begin{table*}[t]
    \caption{NIST SRE 2018 CMN2 results for our submitted system on the fixed condition (EER[\%] / minC / [actC])}
    \label{tab: results1}
    \centering
    \begin{tabular}[c]{clccc}
        \toprule
        \textbf{Front-end} & \textbf{Back-end} & \textbf{Score Norm} & \textbf{Development} & \textbf{Evaluation} \\
        \midrule 
        \multirow{7}*{\tabincell{c}{MFCC i-vector}} & LDA + inW + PLDA & - & 11.00 / 0.681 & 12.21 / 0.745  \\
         & LDA + CORAL + inW + PLDA & - & 11.25 / 0.638 & 12.82 / 0.708 \\
         & LDA + PLDA & AS-Norm2 & 10.64 / 0.603 & \textbf{11.93 / 0.667} \\
         & LDA + CORAL + PLDA & AS-Norm2 & 11.48 / 0.606 & 12.77 / 0.699\\
         & LSDA + CORAL + PLDA & - & 13.27 / 0.624  & 12.89 / 0.727 \\
         & LSDA + CORAL + PLDA & AS-Norm1 & \textbf{12.95 / 0.581} & 12.93 / 0.686 \\
         & LSDA + CORAL + PLDA & AS-Norm2 & 12.18 / 0.590 & 12.64 / 0.685 \\
        \midrule
        \multirow{7}*{\tabincell{c}{DNN tandem \\ i-vector}} & LDA + inW + PLDA & - & 10.64 / 0.669 & 11.32 / 0.691 \\
         & LDA + CORAL + inW + PLDA & - & 09.81 / 0.546 & 10.83 / 0.614  \\
         & LDA + PLDA & AS-Norm1 & 10.04 / 0.549 & 11.00 / 0.603 \\
         & LDA + CORAL + PLDA & AS-Norm1 & 10.15 / 0.524 & 10.97 / 0.603 \\
         & LSDA + CORAL + PLDA & - & 09.03 / 0.533 & 09.91 / 0.606 \\
         & LSDA + CORAL + PLDA & AS-Norm1 & \textbf{08.85 / 0.494} & 09.68 / 0.555 \\
         & LSDA + CORAL + PLDA & AS-Norm2 & 08.10 / 0.495 & \textbf{09.55 / 0.549} \\
        \midrule
        \multirow{4}*{x-vector} & LDA + inW + PLDA & - & 07.77 / 0.587 & 08.89 / 0.587 \\
         & LDA + CORAL + inW + PLDA & - & 07.09 / 0.469 & 07.43 / 0.518 \\
         & LDA + PLDA & AS-Norm2 & 07.17 / 0.479 & \textbf{07.68 / 0.492} \\
         & LDA + CORAL + PLDA & AS-Norm2 & \textbf{07.32 / 0.419} & 07.50 / 0.504 \\
        \midrule
        \multirow{4}*{ResNet v1} & LDA + CORAL + PLDA & - & 07.99 / 0.501 & \textbf{08.18 / 0.540} \\
         & LDA + CORAL + PLDA & AS-Norm1 & \textbf{08.60 / 0.475} & 08.17 / 0.546 \\
         & LDA + CORAL + PLDA & AS-Norm2 & \textbf{08.23 / 0.475} & 08.13 / 0.549 \\
         & cosine similarity & AS-Norm1 & 12.15 / 0.576 & 12.20 / 0.632 \\
        \midrule
        \multirow{4}*{ResNet v2} & LDA + CORAL + PLDA & - & 09.71 / 0.637 & 08.85 / 0.616 \\
         & LDA + CORAL + PLDA & AS-Norm1 & \textbf{09.56 / 0.528} & 08.53 / 0.554 \\
         & LDA + CORAL + PLDA & AS-Norm2 & 08.92 / 0.535 & \textbf{08.19 / 0.546} \\
         & cosine similarity & AS-Norm1 & 13.28 / 0.571 & 13.12 / 0.636 \\
        \midrule
        Fusion &  & & \textbf{5.21 / 0.324 / 0.329} & \textbf{5.53 / 0.392 / 0.409}\\
        \bottomrule
    \end{tabular} 
\end{table*}

\subsubsection{Domain adaptation}
We use Correlation Alignment (CORAL) \cite{sun_return_2016, alam_speaker_2018} to align the distributions of out-of-domain and in-domain features in an unsupervised way by aligning second-order statistics, i.e., covariance. CORAL minimize the distance between the covariance of the out-of-domain and in-domain features, and a linear transformation $\mathbf{A}$ to the source features and the Frobenius norm is used as a matrix distance metric.

In CMN2 condition, we use SRE 18 unlabeled data as the target-domain data and the PLDA training data as the source-domain data, and the linear transformation $\mathbf{A}$ is estimated by CORAL algorithm on these two data.

\subsubsection{In-domain whitening}
To further reduce the variabilities between training data and testing data, in-domain whitening (inW) is applied before PLDA. In-domain whitening calculates the mean and covariance of the in-domain data and applies them to whiten the test data. The whitening transforms are estimated with the SRE 18 unlabeled set and the Speaker In The Wild (SITW) dataset \cite{mclaren_speakers_2016} for CMN2 and VAST respectively.

\subsubsection{Gaussian PLDA}
After whitening, unit-length normalization is applied to the speaker discriminant features. The Gaussian PLDA \cite{garcia-romero_analysis_2011} model with a full covariance residual noise term and a full-rank eigenvoice subspace is then trained for scoring. 

\subsubsection{Cosine similarity}
We use cosine similarity as an alternative back-end for the ResNet based systems. The scores of any given enrolment-test pair are calculated as the cosine similarity of the two features.

\begin{table*}[t]
    \caption{NIST SRE 2018 VAST results for our submitted system on the fixed condition (EER[\%] / minC / [actC])}
    \label{tab: results2}
    \centering
    \begin{tabular}[c]{clccc}
        \toprule
        \textbf{Front-end} & \textbf{Back-end} & \textbf{Score Norm} & \textbf{Development} & \textbf{Evaluation} \\
        \midrule  
        \multirow{2}*{MFCC i-vector} & LDA + PLDA & AS-Norm2 & 11.11 / 0.568 & 17.46 / 0.590 \\
         & LSDA + PLDA & - & 16.05 / 0.597 & 19.90 / 0.613 \\
        \midrule 
        \multirow{2}*{DNN tandem i-vector} & LDA + PLDA & AS-Norm1 & 10.70 / 0.370 & 14.98 / 0.521 \\
         & LSDA + PLDA & - & 10.29 / 0.407 & \textbf{18.41 / 0.515} \\
        \midrule 
        \multirow{1}*{x-vector} & LDA + PLDA & AS-Norm2 & \textbf{08.64 / 0.296} & 13.33 / 0.533 \\
        \midrule 
        \multirow{3}*{ResNet v1} & LDA + PLDA & AS-Norm2 & 08.64 / 0.490 & 13.97 / 0.609 \\
         & cosine similarity & AS-Norm1 & 16.05 / 0.667 & 16.19 / 0.771 \\
         & cosine similarity & AS-Norm2 & 14.81 / 0.630 & 16.83 / 0.739 \\
        \midrule 
        \multirow{1}*{ResNet v2} & LDA + PLDA & - & 13.17 / 0.667 & 16.19 / 0.771 \\
        \midrule 
        Fusion &  &  & \textbf{7.41 / 0.259 / 0.259} & \textbf{12.77 / 0.494 / 0.515} \\
        \bottomrule
    \end{tabular} 
\end{table*}

\begin{table*}[t]
    \caption{CMN2 and VAST results for our post evaluation systems on the fixed condition (EER[\%] / minC / [actC])}
    \label{tab: results3}
    \centering
    \begin{threeparttable}
    \begin{tabular}{clccccc}
    \toprule 
    \multirow{3}*{\textbf{ID}} & \multirow{3}*{\tabincell{c}{\textbf{Encoding Layer}}} & \multirow{3}*{\textbf{Loss}} & \multicolumn{2}{c}{\textbf{CMN2}} & \multicolumn{2}{c}{\textbf{VAST}}\cr
    \cmidrule(lr){4-5} \cmidrule(lr){6-7}
    & & & \textbf{Development} & \textbf{Evaluation} & \textbf{Development} & \textbf{Evaluation} \cr
    \midrule 
    1 & GAP (mean) & Softmax   & 7.85 / 0.501 & 7.43 / 0.557 & 11.93 / 0.486 & 14.93 / 0.542\\ 
    2 & GAP (mean+std) & Softmax   & 7.03 / 0.481 & 7.12 / 0.489 & \textbf{10.29 / 0.333} & 11.75 / 0.480 \\ 
    3 & LDE & Softmax   & \textbf{7.50 / 0.408} & 7.17 / 0.503 & 09.47 / 0.449 & 14.33 / 0.523 \\ 
     \midrule
     4 & GAP (mean) & A-softmax & 6.03 / 0.420 & 6.61 / 0.474 & 10.29 / 0.370 & 13.06 / 0.494 \\ 
     5 & GAP (mean+std) & A-softmax & 5.94 / 0.418 & \textbf{6.14 / 0.463} & \textbf{03.70 / 0.300} & \textbf{10.79 / 0.423 }\\  
     6 & LDE & A-softmax & \textbf{6.03 / 0.354} & \textbf{6.20 / 0.430} & 08.23 / 0.407 & \textbf{13.97 / 0.457}\\
    \midrule
    \multicolumn{3}{l}{Fusion: submitted system} & 5.21 / 0.324 / 0.329 & 5.53 / 0.392 / 0.409  & 7.41 / 0.259 / 0.259 & 12.77 / 0.494 / 0.515 \\
    \multicolumn{3}{l}{Fusion: post-evaluation systems 1-6} & 5.32 / 0.354 / 0.368 & 5.36 / 0.399 / 0.409  & 3.70 / 0.189 / 0.374 & 10.94 / 0.412 / 0.421 \\
    \multicolumn{3}{l}{Fusion: submission + post-evaluation} & \textbf{4.27 / 0.275 / 0.279} & \textbf{4.81 / 0.365 / 0.386} & 0.00 / 0.000 / 0.000 & \textbf{10.74 / 0.401 / 0.406}\\
    \bottomrule
    \end{tabular}
    \end{threeparttable}
\end{table*}

\subsection{Score normalization}
After scoring, all trial results are subject to score normalization. We utilize Adaptive Symmetric Score Normalization (AS-Norm) in our systems. Two variants of AS-Norm associate with the adaptive cohort selection are investigated. The details can be found in ~\cite{sn_analysis_2017}.

The selection of the score normalization corpus is tuned to be optimal for the development data. For CMN2 condition, the score normalization cohort is SRE 18 unlabeled and SRE 16 unlabeled data. For VAST condition, the score normalization cohort is the SITW dataset.

\subsection{System fusion and calibration}
All the subsystems were fused and calibrated using the BOSARIS toolkit \cite{brummer2013bosaris} which learn a scale and a bias for each system by logistic regression approach. The final fusion is an equal-weighted sum of the systems after applying the scale and the bias. Fusion applies to CMN2 and VAST tasks separately.

\section{Submitted system performance}

\subsection{Data preparation}

The training data includes SRE04-16, MIXER 6, Switchboard, VoxCeleb1 \cite{nagrani_voxceleb:_2017} and VoxCeleb2 \cite{chung_voxceleb2:_2018}, resulting 14,467 speakers altogether.

Data augmentation is utilized for both x-vector and  ResNet systems. We adopt the same data augmentation strategy as the Kaldi x-vector recipe. It employs additive noises and reverberation. Reverberation involves convolving room impulse responses (RIR) with audio and the simulated RIRs described by Ko et al. in \cite{ko_study_2017} are used. For additive noise, the MUSAN dataset \cite{musan} is used.

For the MFCC i-vector, DNN Tandem i-vector, and ResNet v1 system, the whole 14,467 speakers are used for training. For the x-vector system, since short-duration utterances and speakers with too little utterances are dropped,  12,459 speakers are used for training. We also train the ResNet v2 system on only the VoxCeleb1 and Voxceleb2 corpus, which includes 7,245 speakers.

\subsection{System performance}

In table \ref{tab: results1}, we present the Equal Error Rate (EER) and SRE18 primary Detection Cost Function (DCF) of our submitted systems in the fixed condition for the CMN2 task. Some observations come from the results. First of all, although the DNN tandem i-vector system utilizes a DNN acoustic model trained on English corpus to extract phonetic aware features, it still shows competitive performance on the language-mismatch testing condition after applying the backend with variabilities compensation, domain adaptation, and score normalization. Also, LSDA obtains approximate 5\% relative performance gain compared to the LDA algorithm. Moreover, although stated in \cite{cai_exploring_2018}, a simple cosine can achieve good performance in the ResNet based systems, the back-end methods including variabilities compensation, domain adaptation, in-domain whitening, and score normalization are beneficial in training-testing mismatch conditions. The last observation is that different combination of the mismatch reduction algorithms can further improve the system performance, and these systems are also complementary to each other.

Table \ref{tab: results2} presents the results of our submitted systems in the fixed condition on VAST task. It is surprising to see that the DNN tandem i-vector system outperform the end-to-end x-vector and ResNet based system in terms of minC on the evaluation set. The reason is that the score normalization setting (AS-Norm2 with 200 cohort scores) used for the x-vector system, which is selected to be optimal for the small trials of the VAST development set (37 utterances, 270 trials), is not well tuned for the evaluation set. Indeed, the minC reaches 0.5 when we use AS-Norm1 with 100 cohort scores on x-vector system.

\section{Post evaluation}
After the evaluation, we further extend our experiments by investigating more encoding layer designs and loss functions for our deep ResNet system.  

First, following the x-vector system that accumulates mean and standard deviation statistics in pooling layer, we boost the GAP layer by adding the global standard deviation statistics of the learned feature maps. Specifically, 256-dimensional mean statistics, as well as 256-dimensional standard deviation statistics, are concatenated together to form the utterance-level representation. Furthermore, we replace the GAP layer with LDE layer. It learns a dictionary with the centers of several clusters and encodes the variable-length inputs into a single utterance-level supervector\cite{cai_exploring_2018, cai_novel_2018}. In our experiments, the number of dictionary components is set to 64.

Second, regarding that the superiority of A-softmax for speaker recognition has been shown in~\cite{cai_exploring_2018}, we use A-softmax loss to replace the basic softmax loss. In our experiment, we use the angular margin $m = 4$. When training the network with A-Softmax loss, we use an annealing optimization strategy, which supervises the network from the original softmax loss gradually to A-softmax loss~\cite{liu_sphereface:_2017}. 
 
Table \ref{tab: results3} shows the post evaluation results of deep ResNet system on NIST SRE 2018 fixed condition. The best back-end setting and score normalization for the CMN2 systems are tuned on the development set and then apply to the evaluation set, while the VAST system is tuned directly on the evaluation set since the development set for VAST is small, and the results on such small development set are unreliable to some extent.

From table \ref{tab: results3}, we can see that the LDE layer shows great potential in front-end modeling under language-mismatch condition. Also, GAP layer integrated with standard deviation statistics obtains better performance than the basic GAP layer in both CMN2 and VAST tasks and achieves the best performance in the VAST task. As demonstrated in~\cite{cai_exploring_2018, liu_sphereface:_2017}, the system with A-softmax loss always shows an apparent reduction in both EER and DCF compared to the system with softmax loss.

We also provide the fusion results of our post-evaluation systems in table \ref{tab: results3}. The final fusion with submitted systems and post-evaluation systems obtains 17.8\%, 7.4\%, and 23.2\% relative improvements in terms of minC over the CMN2 development, CMN2 evaluation, and VAST evaluation respectively. 

\section{Conclusions}
In this paper, our submitted DKU-SMIIP system for the NIST SRE 2018 is described. Various kinds of state-of-the-art speaker embedding extractors are explored. We also utilize variabilities compensation, domain adaptation, in-domain whitening, and score normalization algorithms to reduce the mismatch condition between training and testing data. The experimental results show the potential of deep ResNet for large-scale TISV and demonstrate the significance of LDE layer and A-softmax loss. 

\bibliographystyle{IEEEtran}
\bibliography{mybib}

\end{document}